\documentclass[11pt,notitlepage,nofootinbib]{revtex4-1}

\usepackage{epsf}
\usepackage{latexsym}
\usepackage{amssymb}
\usepackage{amsmath}
\usepackage{bm}
\usepackage[dvips]{graphicx}
\usepackage{array}

\def\d3{^{(3)}\nabla}

\begin{document}

\title{Secondary CMB anisotropies from bulk motions in the presence of stochastic magnetic fields}

\author{Kerstin E. Kunze}

\email{kkunze@usal.es}

\affiliation{Departamento de F\'\i sica Fundamental and {\sl IUFFyM}, Universidad de Salamanca,
 Plaza de la Merced s/n, 37008 Salamanca, Spain}

\begin{abstract}
Bulk motions of electrons along the line of sight induce secondary temperature fluctuations in the 
post-decoupling, reionized universe.
In the presence of a magnetic field not only the scalar mode but also the vector mode act
as a source for the bulk motion. The resulting angular power spectrum of temperature 
anisotropies of the cosmic microwave background is calculated assuming a simple model
of reionization.
Contributions from the standard adiabatic, curvature mode
and a non helical magnetic field are included. The contribution due to magnetic fields with  field strengths 
of order nG and negative magnetic  spectral indices  becomes important for multipoles larger than
$\ell \sim  10^4$.

\end{abstract}

\maketitle

\section{Introduction}
\setcounter{equation}{0}

Current measurements of the cosmic microwave background (CMB) give a very precise
image of the CMB anisotropies. The acoustic peaks have been detected and measurements are now moving
to larger multipoles with the {\it South Pole Telescope} (SPT) covering $2000<\ell<9400$ \cite{spt}
and the {\it Atacama Cosmology Telescope} (ACT) measuring   $500<\ell<10000$ \cite{act}.
Primary CMB anisotropies are calculated in linear perturbation theory. 
The observed peak structure as well as the damping tail for $500<\ell<3000$ are, in general, well 
explained by the $\Lambda$CDM model \cite{planck,wmap9,acbar,quad}. At higher multipoles secondary CMB
anisotropies caused by nonlinear effects can dominate over the primary signal.
After decoupling at $z_{dec}=1088$ \cite{wmap9} scattering of CMB photons off free electrons
becomes important again within the reionized universe. 
From quasar absorption spectra it is known that the universe is completely reionized
at redshifts $z\sim 6$ \cite{quasar}.
Within the $\Lambda$CDM model CMB anisotropies rule out  reionization  before redshifts of  around $z\sim10$ \cite{wmap9}.
Moving ionized matter induces temperature fluctuations. At linear order this is described by the 
Doppler term in the final line-of-sight integral of the brightness perturbation which however on small scales
becomes very small. This is due to rapid oscillations in the integral averaging to zero assuming a
homogeneous electron distribution. 
However, taking into account the perturbations in the baryon energy density  this is no longer the case and 
at second order there can be an important contribution which is also known as the Ostriker-Vishniac effect \cite{vishniac,ov,jk,hu,hw1}.
It has been shown that the Ostriker-Vishniac effect is the largest second order contribution due to electron photon interaction \cite{HuSS,dj}.
Thus primary fluctuations on small scales are erased once the universe becomes reionized but 
secondary ones are generated.

Magnetic fields present before decoupling have an effect on the CMB anisotropies and in difference  to 
the standard $\Lambda$CDM model they also source vector modes \cite{mkk,yikm,pfp,sl,kk1}.
The field strength of a putative magnetic field present before decoupling has been limited by the 
{\it Planck} data in combination with ACT and SPT data to be less than 3.4 nG  with a preference for negative spectral indices
\cite{planck}. 
The Ostriker-Vishniac effect is determined by the baryon density perturbation and the baryon velocity perturbation. Whereas in the case of the 
$\Lambda$CDM model the velocity field  is purely irrotational, there is in addition  a vortical component in the 
presence of a magnetic field. In the following the Ostriker-Vishniac effect is calculated in the presence of a stochastic magnetic field using the
analytical treatment of \cite{hu,hw1}.

\section{Secondary anisotropies}
\setcounter{equation}{0}

The motion of the electrons and ions induce temperature fluctuations in the CMB determined by \cite{hu}
\begin{eqnarray}
\Theta(\bm{\hat{n}})=\int dDg(D)\bm{\hat{n}}\cdot\bm{V}_b(\bm{x}),
\label{e1}
\end{eqnarray}
where $\bm{\hat{n}}$ is the direction in the sky and $\bm{V}_b$ is the baryon velocity along the line of sight at
$\bm{x}=D\bm{\hat{n}}$. Moreover, $D(z)$ is the conformal distance or look-back time from the observer 
at $z=0$ given by \cite{hu}
\begin{eqnarray}
D(z)=\int_0^z\frac{H_0}{H(z')}dz'
\end{eqnarray}
and $H_0$ is the Hubble parameter today. This is related to conformal time $\eta$ by $D(\eta)=a_0H_0(\eta_0-\eta)$. 
$g$ is the visibility function which determines the probability of scattering of a CMB photon. At linear order $g$ is just a function of time. 
Fluctuations in the baryon energy density along the line of sight change the number density of potential scatteres for the CMB photons
and thus change their scattering probability. Therefore with $g\rightarrow g(\eta)+\delta g(\bm{x},\eta)$ the velocity field in equation (\ref{e1}) can be
written in terms of an effective velocity $\delta \bm{V}_b$ as $g\bm{V}_b\rightarrow g(\bm{V}_b+\delta\bm{V}_b)= g(1+\frac{\delta g}{g})\bm{V}_b$.
Thus the bulk motion of the scatteres with inhomogeneous number densities effectively contributes a velocity perturbation 
$\delta\bm{V}_b(\bm{x},\eta)$ at second order given by 
\begin{eqnarray}
\delta\bm{V}_b(\bm{x},\eta)=\Delta_b(\bm{x},\eta)\bm{V}_b(\bm{x},\eta).
\label{e2}
\end{eqnarray}
Expanding equation (\ref{e2}) in terms of spherical harmonics yields to (e.g. \cite{ks}),
\begin{eqnarray}
\delta\bm{V}_{b,i}(\bm{x},\eta)=\sum_{m=0,\pm 1}\sum_{\bm{k}}\delta V_b^{(m)}(\bm{k},\eta)Q_i^{(m)}(\bm{k},\bm{x}),
\label{e03}
\end{eqnarray}
where $Q_i^{(0)}=-k^{-1}\nabla_i Q^{(0)}$, $Q^{(0)}=e^{i\bm{k}\cdot\bm{x}}$ in a flat universe, so that $Q_i^{(0)}=-i\frac{k_i}{k}e^{i\bm{k}\cdot\bm{x}}$.
A coordinate system with basis vectors $\bm{\hat{e}}_{\bm{k}}^{(i)}$ is chosen such that  $\bm{\hat{e}}_{\bm{k}}^{(3)}||\bm{k}$. 
Moreover, the helicity basis is defined by $\bm{\hat{e}}_{\bm{k}}^{(\pm 1)}=-\frac{i}{\sqrt{2}}\left(\bm{\hat{e}}^{(1)}_{\bm{k}}\pm i\bm{\hat{e}}^{(2)}_{\bm{k}}\right)$.
Finally, $Q_i^{(\pm 1)}=\left(\bm{\hat{e}}_{\bm{k}}^{(\pm 1)}\right)_ie^{i\bm{k}\cdot\bm{x}}$.
Following \cite{hw} the brightness perturbation $\Theta(\bm{x},\bm{\hat{n}},\eta)$ including 
only scalar ($m=0$) and  vector ($m=\pm 1$) modes 
is given by
\begin{eqnarray}
\Theta(\bm{x},\bm{\hat{n}},\eta)=\int\frac{d^3k}{(2\pi)^3}\sum_{\ell}\sum_{m=-1}^{m=1}\Theta_{\ell}^{(m)}(\bm{k},\eta)G_{\ell}^m(\bm{k},\bm{x}),
\end{eqnarray}
where $G_{\ell}^m=(-i)^{\ell}\sqrt{\frac{4\pi}{2\ell+1}}Y^m_{\ell}(\bm{\hat{n}})e^{i\bm{k}\cdot\bm{x}}$.
Using that \cite{hw} $n^iQ_i^{(0)}=G^0_1$ and $n^iQ_i^{(\pm 1)}=G_1^{\pm 1}$ then
\begin{eqnarray}
\hat{n}\cdot\delta{\bm{V}}_b(\bm{x},\eta)&=&\sum_{\bm{k}}\left[\delta V_b^{(0)}(\bm{k},\eta)G_1^{(0)}+
\delta V_b^{(\pm 1)}(\bm{k},\eta)G_1^{(\pm 1)}\right].
\end{eqnarray}
In the continuum limit $\sum_{\bm{k}}\rightarrow\int\frac{d^3k}{(2\pi)^3}$ equations (\ref{e2}) and (\ref{e03}) yield,
\begin{eqnarray}
\delta V_b^{(0)}(\bm{k},\eta)&=&\int\frac{d^3k_1}{(2\pi)^3}\left[V_b^{(0)}(\bm{k}_1,\eta)\bm{\hat{e}}^{(3)}_{\bm{k}_1}\cdot\bm{\hat{e}}_{\bm{k}}^{(3)}
+iV_b^{(+1)}(\bm{k}_1,\eta)\bm{\hat{e}}_{\bm{k}_1}^{(+1)}\cdot
\bm{\hat{e}}_{\bm{k}}^{(3)}
\right.\nonumber\\
&&\left.+iV_b^{(-1)}(\bm{k}_1,\eta)\bm{\hat{e}}_{\bm{k}_1}^{(-1)}\cdot\bm{\hat{e}}^{(3)}_{\bm{k}}\right]\Delta_b(\bm{k}-\bm{k}_1,\eta)\\
\delta V_b^{(+1)}(\bm{k},\eta)&=&-\int\frac{d^3k_1}{(2\pi)^3}\left[-iV_b^{(0)}(\bm{k}_1,\eta)\bm{\hat{e}}^{(3)}_{\bm{k}_1}\cdot\bm{\hat{e}}_{\bm{k}}^{(-1)}
+V_b^{(+1)}(\bm{k}_1,\eta)\bm{\hat{e}}^{(+1)}_{\bm{k}_1}\cdot\bm{\hat{e}}^{(-1)}_{\bm{k}}
\right.\nonumber\\
&&\left.
+V_b^{(-1)}(\bm{k}_1,\eta)\bm{\hat{e}}_{\bm{k}_1}^{(-1)}
\cdot\bm{\hat{e}}^{(-1)}_{\bm{k}}\right]\Delta_b(\bm{k}-\bm{k}_1,\eta)
\label{dv1}\\
\delta V_b^{(-1)}(\bm{k},\eta)&=&-\int\frac{d^3k_1}{(2\pi)^3}\left[-iV_b^{(0)}(\bm{k}_1,\eta)\bm{\hat{e}}^{(3)}_{\bm{k}_1}\cdot\bm{\hat{e}}_{\bm{k}}^{(+1)}
+V_b^{(+1)}(\bm{k}_1,\eta)\bm{\hat{e}}^{(+1)}_{\bm{k}_1}\cdot\bm{\hat{e}}^{(+1)}_{\bm{k}}
\right.\nonumber\\
&&\left.
+V_b^{(-1)}(\bm{k}_1,\eta)\bm{\hat{e}}_{\bm{k}_1}^{(-1)}
\cdot\bm{\hat{e}}^{(+1)}_{\bm{k}}\right]\Delta_b(\bm{k}-\bm{k}_1,\eta).
\label{dv-1}
\end{eqnarray}

Finally, using the expressions for the line of sight integral for the Doppler term \cite{hw} for the effective velocity perturbation $\delta V_b^{(m)}(\bm{k},\eta)$ results for the  scalar mode contribution in
\begin{eqnarray}
\frac{\Theta_{\ell}^{(0)}(\bm{k}, \eta_0)}{2\ell +1}=\int_0^{\eta_0}d\eta g(\eta)\delta V_b^{(0)}(\bm{k},\eta)j_{\ell}^{(1 0)}\left[k(\eta_0-\eta)\right].
\end{eqnarray}
The vector mode contribution is determined by
\begin{eqnarray}
\frac{\Theta_{\ell}^{(\pm 1)}(\bm{k},\eta_0)}{2\ell +1}=\int_0^{\eta_0}d\eta g(\eta)\delta V_b^{(\pm 1)}(\bm{k},\eta)j_{\ell}^{(1 1)}\left[k(\eta_0-\eta)\right].
\end{eqnarray}
The radial functions are given by
\begin{eqnarray}
j_{\ell}^{(1 0)}(x)&=&j_{\ell}'(x),\nonumber\\
j_{\ell}^{(1 1)}(x)&=&\sqrt{\frac{\ell(\ell+1)}{2}}\frac{j_{\ell}(x)}{x}.
\end{eqnarray}
For large $\ell$ and small scales the contribution due to the scalar mode can be neglected as the integrand is an oscillating function leaving
a negligible effect \cite{vishniac,ov,jk,hu,hw1}.
Therefore, the angular power spectrum of the temperature anisotropies due to the vector mode is found to be
\begin{eqnarray}
C_{\ell}^{(V)}=\frac{2}{\pi}\int\frac{dk}{k}k^3(2\ell+1)^{-2}
\langle\Theta_{\ell}^{(+1)\;*}(\bm{k},\eta_0)\Theta^{(+1)}_{\ell}(\bm{k},\eta_0)
+\Theta^{(-1)\;*}_{\ell}(\bm{k},\eta_0) \Theta^{(-1)}_{\ell}(\bm{k},\eta_0)\rangle,
\label{cl}
\end{eqnarray}
where 
\begin{eqnarray}
&(2\ell+1)^{-2}&\langle\Theta_{\ell}^{(+1)*}(\bm{k},\eta_0)\Theta^{(+1)}_{\ell}(\bm{k},\eta_0)+\Theta_{\ell}^{(-1)*}(\bm{k},\eta_0)\Theta^{(-1)}(\bm{k},\eta_0)
\rangle\nonumber\\
&=&\int_0^{\eta_0}d\eta g(\eta)\int_0^{\eta_0}d\eta' g(\eta')\frac{\ell(\ell+1)}{2}
\frac{j_{\ell}\left[k(\eta_0-\eta)\right]}{k(\eta_0-\eta)}\frac{j_{\ell}\left[k(\eta_0-\eta')\right]}{k(\eta_0-\eta')}\nonumber\\
&\times&
\langle\delta V_b^{(+1)*}(\bm{k},\eta')\delta V_b^{(+1)}(\bm{k},\eta)
+\delta V_b^{(-1)*}(\bm{k},\eta')\delta V_b^{(-1)}(\bm{k},\eta)\rangle.
\label{thethe}
\end{eqnarray}

In order to calculate the two point function of $\delta V_b^{(\pm 1)}$ the solutions for the baryon velocity fields of the scalar and vector modes 
at linear order are required which will be given in the following. In the baryon density perturbation as well as the baryon velocity at linear order are included
the contributions from the adiabatic, primordial curvature mode as well as the magnetic mode.
The magnetic field is assumed to be a nonhelical, gaussian random field with a two point function in $k$-space given by
\begin{eqnarray}
\langle B_i(\bm{k})B_j(\bm{k'})\rangle=\delta_{\bm{k},\bm{k'}}P_B(k)\left(\delta_{ij}-\frac{k_ik_j}{k^2}\right),
\end{eqnarray}
where the spectrum is chosen to be of the form \cite{kk1}
\begin{eqnarray}
P_B(k,k_m,k_L)=A_B\left(\frac{k}{k_L}\right)^{n_B}W(k,k_m)
\end{eqnarray}
where $k_L$ is a pivot wave number chosen to be 1 Mpc$^{-1}$ and  $W(k,k_m)=\pi^{-3/2}k_m^{-3}e^{-(k/k_m)^2}$
is a gaussian window function. $k_m$ corresponds to the largest scale
damped due to radiative viscosity before decoupling \cite{sb,jko}. $k_m$ has its largest value at recombination 
\begin{eqnarray}
k_m=286.91\left(\frac{B}{\rm nG}\right)^{-1}{\rm Mpc}^{-1}
\end{eqnarray}
for the bestfit parameters of {\it WMAP 9}  data only \cite{kuko,wmap9}.

\subsection{Scalar mode}

The perturbation equations of baryons and cold dark matter (e.g. \cite{ks,sl,kk1})
can be combined to yield the evolution equation of the total matter  perturbation 
$\Delta_m=\tilde{R}_c\Delta_c+\tilde{R}_b\Delta_b$, where $\tilde{R}_i=\frac{\rho_i}{\rho_m}$
denotes the fractional energy density of the component $i=b,c$ w.r.t.  the total matter density $\rho_m=\rho_c+\rho_b$. Namely,
for the magnetic mode during matter domination it evolves as
\begin{eqnarray}
\frac{d^2\Delta_m}{dw^2}+\frac{2}{w}\frac{d\Delta_m}{dw}-\frac{6}{w^2}\Delta_m=-\frac{1}{3}\frac{\Omega_{\gamma,0}}{\Omega_{m,0}}
\left(\frac{w_0}{w}\right)^2L(\bm{k})
\label{e3}
\end{eqnarray}
where $w\equiv k\eta$ and $w_0\equiv k\eta_0$.  $L(\bm{k})$ is related to the Lorentz force $\bm{L}(\bm{x},\eta)=\left[(\bm{\nabla}\times\bm{B})\times
\bm{B}\right](\bm{x},\eta)$ (e.g. \cite{kk1}). 
Similary for the baryon density perturbation,
\begin{eqnarray}
\frac{d^2\Delta_b}{dw^2}+\frac{2}{w}\frac{d\Delta_b}{dw}=\frac{6}{w^2}\Delta_m-\frac{1}{3}\frac{\Omega_{\gamma,0}}{\Omega_{b,0}}
\left(\frac{w_0}{w}\right)^2L(\bm{k})
\label{e4}
\end{eqnarray}
Solving equation (\ref{e3}) and using the solution for  $\Delta_m$ in equation (\ref{e4}) shows that the amplitude of the growing mode of the 
baryon perturbation is the same as that of the total matter perturbation. Thus also in the magnetized case the baryon 
density perturbation is following the total matter perturbation \cite{sesu}. In particular, the growing mode of the baryon perturbation due to the magnetic mode $(B)$ is given by
\begin{eqnarray}
\Delta_b^{(B)}(k,\eta)=D(\eta)\Delta_b^{(B)}(k,\eta_0)
\end{eqnarray}
where the growth factor 
\begin{eqnarray}
D(\eta)=\left(\frac{\eta}{\eta_0}\right)^2
\end{eqnarray}
has been introduced and
\begin{eqnarray}
\Delta_b^{(B)}(k,\eta_0)=-\frac{L}{30}\frac{\Omega_{\gamma,0}}{\Omega_{m,0}}(k\eta_0)^2\left(\frac{\eta_0}{\eta_i}\right)^2
\label{e5}
\end{eqnarray}
where $\eta_i$ is some initial time at which $\Delta_b^{(B)}(x_i)=0$. Neglecting perturbations from before  decoupling
the initial time is set to $\eta_i=\eta_{dec}$  \cite{kor,sesu}.  
Matter perturbations in a magnetized medium cannot grow on scales below the magnetic Jeans length as the magnetic pressure will prevent any 
further collapse \cite{kor,sesu}. Therefore the simplest approach is to assume $D(\eta)=0$ on scales corresponding to 
$k>k_{J}$ where $k_J$ is the wave number corresponding to the magnetic Jeans scale. It is given by \cite{sesu}
\begin{eqnarray}
\left(\frac{k_J}{{\rm Mpc}^{-1}}\right)=\left[14.8\left(\frac{\Omega_m}{0.3}\right)^{\frac{1}{2}}\left(\frac{h}{0.7}\right)
\left(\frac{B}{10^{-9}{\rm G}}\right)^{-1}\left(\frac{k_L}{{\rm Mpc}^{-1}}\right)^{\frac{n_B+3}{2}}\right]^{\frac{2}{n_B+5}}.
\end{eqnarray}

In the case of the adiabatic, primordial curvature mode $(ad)$ the baryon density perturbation follows the total matter perturbation which is
given by (e.g. \cite{hu,hw1})
\begin{eqnarray}
\Delta_m^{(ad)}(\bm{k},\eta)=D(\eta)\Delta_m^{(ad)}(\bm{k},\eta_0).
\end{eqnarray}

The linear matter power spectrum $P_m$  is defined by
\begin{eqnarray}
\langle\Delta_m^*(\bm{k},\eta_0)\Delta_m(\bm{k}',\eta_0)\rangle=P_m(k)\delta_{\bm{k},\bm{k'}}.
\end{eqnarray}
For the adiabatic, curvature mode it is given by
\begin{eqnarray}
P^{(ad)}_m(k)=\frac{2\pi^2}{k^3}\left(\frac{k}{a_0H_0}\right)^4\frac{4}{25}A_s\left(\frac{k}{k_p}\right)^{n_s-1}T^2(k),
\end{eqnarray}
where the transfer function $T(k)$ is given by \cite{pu,bbks}
\begin{eqnarray}
T(k)=\frac{\ln(1+2.34q)}{2.34q}\left[1+3.89q+(16.1q)^2+(5.46q)^3+(6.71q)^4\right]^{-\frac{1}{4}}
\end{eqnarray}
where $q=\frac{k}{\Omega_{m,0}h^2 {\rm Mpc}^{-1}}$.

For the magnetic mode the matter power spectrum is found to be from equation (\ref{e5})
\begin{eqnarray}
P_m^{(B)}(k)=\frac{2\pi^2}{k^3}\left(\frac{k}{a_0H_0}\right)^4\frac{4}{225}(1+z_{dec})^2\left(\frac{\Omega_{\gamma,0}}{\Omega_{m,0}}\right)^2
{\cal P}_L(k),
\end{eqnarray}
where ${\cal P}_L(k)$ is the dimensionless power spectrum determining the two point function of the Lorentz term
$\langle L^*(\bm{k})L(\bm{k'})\rangle=\frac{2\pi^2}{k^3}{\cal P}_L(k)$ given by \cite{kk1}
\begin{eqnarray}
{\cal P}_L(k)&=&\frac{9}{\left[\Gamma\left(\frac{n_B+3}{2}\right)\right]^2}\left(\frac{\rho_{B,0}}{\rho_{\gamma,0}}\right)^2
\left(\frac{k}{k_m}\right)^{2(n_B+3)}e^{-\left(\frac{k}{k_m}\right)^2}\nonumber\\
&\times&\int_0^{\infty}dz z^{n_B+2}e^{-2\left(\frac{k}{k_m}\right)^2z^2}\int_{-1}^1 dxe^{2\left(\frac{k}{k_m}\right)^2zx}
(1-2zx+z^2)^{\frac{n_B-2}{2}}\nonumber\\
&\times&\left[1+2z^2+(1-4z^2)x^2-4zx^3+4z^2x^4\right],
\label{pL}
\end{eqnarray}
and $x\equiv\frac{\bm{k}\cdot\bm{q}}{kq}$ and $z\equiv\frac{q}{k}$ where $\bm{q}$ is the wave number over which the 
resulting convolution integral is calculated.

During the matter dominated era the baryon velocity is determined by (e.g. \cite{ks})
\begin{eqnarray}
V_b=-k^{-1}\dot{\Delta}_b.
\end{eqnarray}
Therefore the total baryon density perturbation and the baryon velocity at linear order  are found to be 
\begin{eqnarray}
\Delta_b(\bm{k},\eta)&=&D(\eta)\Delta_b^{(ad)}(\bm{k},\eta_0)+D(\eta)\Delta_b^{(B)}(\bm{k},\eta_0)\\
V^{(0)}_b(\bm{k},\eta)&=&-\frac{\dot{D}(\eta)}{k}\Delta_b^{(ad)}(\bm{k},\eta_0)-\frac{\dot{D}(\eta)}{k}\Delta_b^{(B)}(\bm{k},\eta_0).
\end{eqnarray}
For simplicity, it is assumed that  there is no cross correlation between the adiabatic, curvature mode and the magnetic mode.

\subsection{Vector mode}

After decoupling and assuming matter domination the baryon vorticity $V_b^{(\pm 1)}$ is determined by  (e.g. \cite{kk2})
\begin{eqnarray}
\frac{dV_b^{(\pm 1)}}{dw}+\frac{2}{w}V_b^{(\pm 1)}=-\frac{1}{6}\frac{\Omega_{\gamma,0}}{\Omega_{b,0}}\pi_B^{(\pm 1)}\left(\frac{w_0}{w}\right)^2,
\end{eqnarray}
where as before $w=k\eta$ and $w_0=k\eta_0$.
The dominant solution is  sourced by the magnetic field and is given by,
\begin{eqnarray}
V_b^{(\pm 1)}(\bm{k},\eta)=F_k(\eta)\pi_B^{(\pm 1)}(\bm{k})
\end{eqnarray}
where the growth factor is given by
\begin{eqnarray}
F_k(\eta)=-\frac{1}{6}\frac{\Omega_{\gamma,0}}{\Omega_{b,0}}k\eta_0\left(\frac{\eta_0}{\eta}\right). 
\end{eqnarray}
The two point function of the anisotropic stress term $\pi^{(\pm 1)}(\bm{k})$ is given by (e.g. \cite{kk2})
\begin{eqnarray}
\langle\pi_B^{(+1)*}(\bm{k})\pi_B^{(+1)}(\bm{k}')+\pi_B^{(-1)*}(\bm{k})\pi_B^{(-1)}(\bm{k'})\rangle
=\frac{2\pi^2}{k^3}{\cal P}_{\langle\pi^{(\pm 1)*}_B\pi^{(\pm 1)}_B\rangle}(k)\delta_{\bm{k},\bm{k}'}
\end{eqnarray}
where for a non helical magnetic field
\begin{eqnarray}
{\cal P}_{\langle\pi^{(\pm 1)*}_B\pi^{(\pm 1)}_B\rangle}(k)&=&\frac{72}{\left[\Gamma\left(\frac{n_B+3}{2}\right)\right]^2}
\left(\frac{\rho_{B,0}}{\rho_{\gamma,0}}\right)^2\left(\frac{k}{k_m}\right)^{2(3+n_B)}e^{-\left(\frac{k}{k_m}\right)^2}
\int_0^{\infty}dz z^{n_B+2}e^{-2\left(\frac{k}{k_m}\right)^2z^2}\nonumber\\
&\times&\int_{-1}^1dx e^{2\left(\frac{k}{k_m}\right)^2zx}(1-2zx+z^2)^{\frac{n_B-2}{2}}(1-x^2)(1+z^2-3zx+2z^2x^2),
\end{eqnarray}
and $x$ and $z$  are defined as in the case of the scalar mode (cf. equation (\ref{pL})).

\subsection{Results}

The expression for the angular power spectrum of the secondary  CMB anisotropies $C_{\ell}^{(V)}$ (\ref{cl}) 
involves the two point function of $\delta V_b^{(\pm 1)}$ (cf. equations (\ref{dv1}) and (\ref{dv-1}))
which is found to be 
\begin{eqnarray}
&\,&\langle\delta V_b^{(+1)*}(\bm{k},\eta')\delta V_b^{(+1)}(\bm{k},\eta)+\delta V_b^{(-1)*}(\bm{k},\eta')\delta V_b^{(-1)}(\bm{k},\eta)\rangle
=
\frac{1}{4}\int_0^{\infty}dy_1\int_{-1}^1 dy D(\eta)D(\eta')
\nonumber\\
&&\times
\left[\frac{k}{\pi^2}\dot{D}(\eta)\dot{D}(\eta')\frac{(1-y^2)(1-2yy_1)}{1-2yy_1+y_1^2}
\left[P_m^{(ad)}(k_1)P_m^{(ad)}(|\bm{k}-\bm{k}_1|)+P_m^{(ad)}(k_1)P_m^{(B)}(|\bm{k}-\bm{k}_1|)
\right.
\right.
\nonumber\\
&&
\left.\left.\;\;
+P_m^{(B)}(k_1)P_m^{(ad)}(|\bm{k}-\bm{k}_1|)+P_m^{(B)}(k_1)P_m^{(B)}(|\bm{k}-\bm{k}_1|)
\right]
\right.
\nonumber\\
&&
\left.\;\;
+F_k(\eta)F_k(\eta')\frac{1+y^2}{y_1}{\cal P}_{\langle\pi_B^{(\pm 1)*}\pi_B^{\pm 1}\rangle}(k_1)
\left[P_m^{(ad)}(|\bm{k}-\bm{k}_1|)+P_m^{(B)}(|\bm{k}-\bm{k}_1|)\right]
\right],
\end{eqnarray}
where $y_1\equiv\frac{k_1}{k}$ and $y\equiv\frac{\bm{k}\cdot\bm{k}_1}{kk_1}$.
Since this expression is separable in the conformal times $\eta$ and $\eta'$ equation (\ref{thethe}) can be written as 
\begin{eqnarray}
&(&2\ell+1)^{-2}\langle\Theta_{\ell}^{(+1)*}(\bm{k},\eta_0)\Theta^{(+1)}_{\ell}(\bm{k},\eta_0)+\Theta_{\ell}^{(-1)*}(\bm{k},\eta_0)\Theta^{(-1)}(\bm{k},\eta_0)
\rangle
\nonumber\\
&&=\frac{\ell(\ell+1)}{2}\sum_{i=1}^2\beta_i(k)
U_{i,\ell}(k,\eta_0)^2
\label{e239}
\end{eqnarray}
where
\begin{eqnarray}
U_{i,\ell}(k,\eta_0)=\int_0^{\eta_0} d\eta g(\eta)\alpha_i(\eta,k)\frac{j_{\ell}\left[k(\eta_0-\eta)\right]}{k\left(\eta_0-\eta\right)}.
\end{eqnarray}
Moreover, $\alpha_1(\eta, k)=D(\eta)\dot{D}(\eta)$ and $\alpha_2(\eta,k)=D(\eta)F_k(\eta)$ and the remaining terms in 
equation (\ref{e239}) are collected in $\beta_1(k)$ and $\beta_2(k)$, respectively.
As shown in \cite{hw1}  $U_{i,\ell}$ can be approximated by
\begin{eqnarray}
U_{i,\ell}(k,\eta_0)\simeq\sqrt{\frac{\pi}{2\ell}}\frac{g(\eta_{\ell})\alpha_i(\eta_{\ell},k)}{k^2(\eta_0-\eta_{\ell})},
\hspace{2cm}
\eta_{\ell}=\eta_0-\frac{\ell+\frac{1}{2}}{k}.
\end{eqnarray}  

\begin{figure}[h!]
\centerline{\epsfxsize=3in\epsfbox{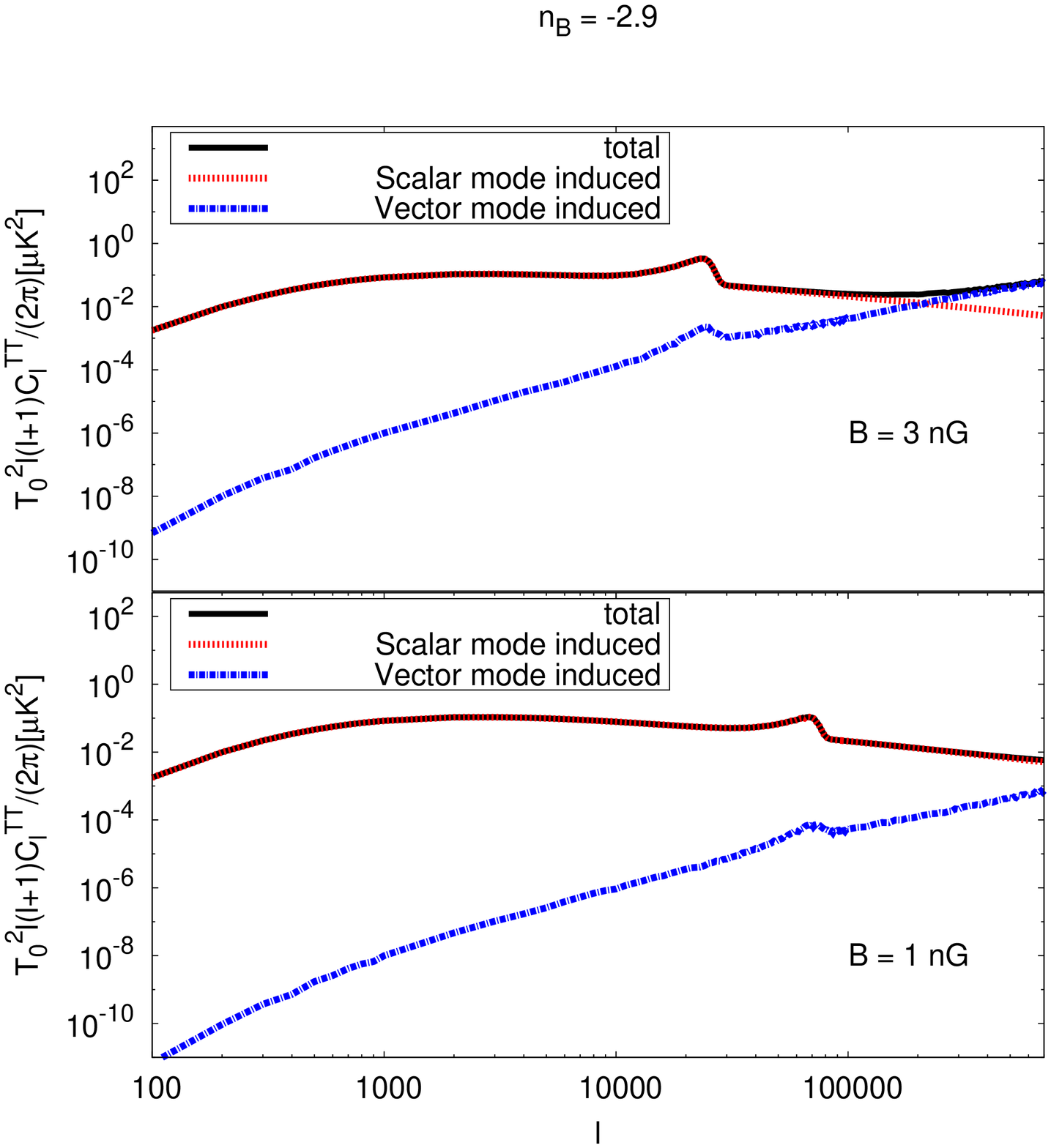}
\hspace{0.1cm}
\epsfxsize=3in\epsfbox{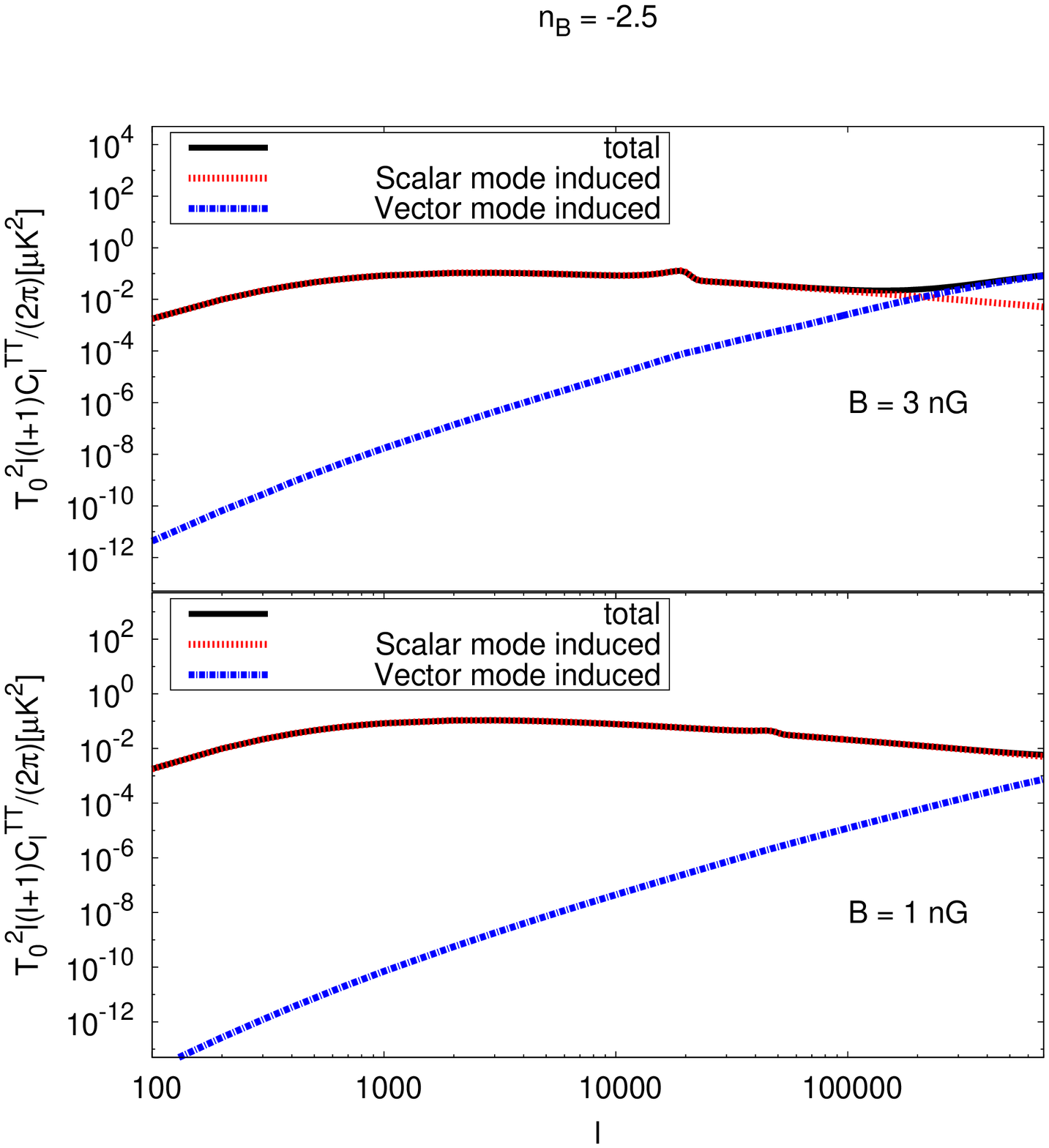}
}
\caption{Contributions and total angular power spectrum of the temperature fluctuations of the Ostriker-Vishniac effect 
for magnetic field strengths $B=1$ nG and $B=3$ nG for spectral indices $n_B=-2.9$ 
({\it left panel}) and $n_B=-2.5$ ({\it right panel}) .
The contributions induced by the scalar mode and the one of the vector mode are shown together with the total 
amplitude of the angular power spectrum.}
\label{fig1}
\end{figure}

\begin{figure}[h!]
\centerline{\epsfxsize=3in\epsfbox{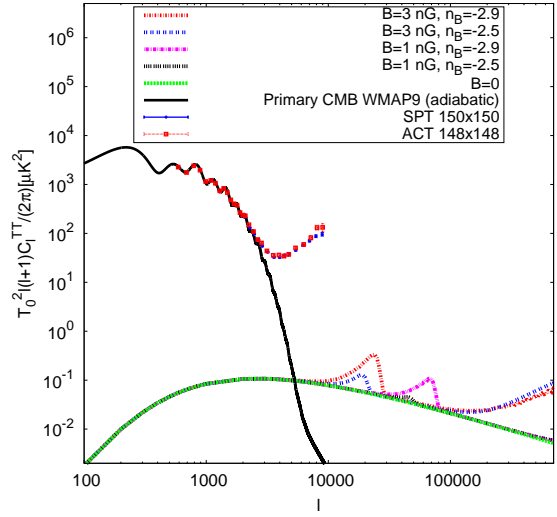}}
\caption{The  total angular power spectrum of the temperature fluctuations of the Ostriker-Vishniac effect   for different values of the magnetic field strength and  spectral indices. For comparison the secondary CMB anisotropies induced by the adiabatic mode is included.
Also shown are the SPT data at 150 GHz, ACT data at 148 GHz  and the primary CMB anisotropies for the {\it WMAP 9}  best fit parameters.}
\label{fig2}
\end{figure}

We are interested in the CMB anisotropies generated by the bulk motion in the post decoupling universe.
After a long period after decoupling at around $z_{dec}=1088$  with a small residual fraction of matter in an ionized state the universe is reionized at 
some redshift $z_r=10.6$ as indicated, e.g., by {\it WMAP 9} \cite{wmap9}.
The amplitude of the temperature fluctuations depends on the visibility function $g(\eta)$ and hence on the reionization history of the universe.
For simplicity it is assumed that the visibility function can be approximated by a gaussian as  \cite{gs,jk}
\begin{eqnarray}
g(\eta)=\frac{1-\exp(-\tau_{\rm r})}{\sqrt{\pi}\Delta\eta_{\rm r}}\exp\left[-\left(\frac{\eta-\eta_r}{\Delta\eta_r}\right)^2\right]
\end{eqnarray}
where the optical depth to the epoch of reionization at $\eta_r$ is  $\tau_r=0.089$ from {\it WMAP 9} data only \cite{wmap9}.
Moreover, following \cite{gs}  the width of the re-scattering surface is chosen to be determined by
$\Delta\eta_r=0.25\eta_r$.
This yields to the following expressions for the angular power spectrum.
The secondary CMB temperature anisotropies  sourced by  the scalar mode at linear order are determined by, 
\begin{eqnarray}
C_{\ell}^{V,S}&=&2\pi\frac{\ell+1}{(\ell+\frac{1}{2})^2}\left(1-e^{-\tau_r}\right)^2(1+z_r)\int_0^{\infty}\frac{dk}{k}\left(\frac{k}{a_0H_0}\right)^4
\kappa_{\ell}^6e^{-32\left[\sqrt{1+z_r}\kappa_{\ell}-1\right]^2}\int_0^{\infty}dy_1
\nonumber\\
&\times&
\int_{-1}^{1}dy(1-y^2)(1-2yy_1)\vartheta^{-2}\left[\frac{16}{625}A_s^2\left(\frac{k}{k_p}\right)^{2(n_s-1)}y_1^{n_s}\vartheta^{n_s}
T^2(ky_1)T^2(k\vartheta)
\right.
\nonumber\\
&+&\left.
\frac{16}{5625}A_s(1+z_{dec})^2\left(\frac{\Omega_{\gamma,0}}{\Omega_{m,0}}\right)^2\left(\frac{k}{k_p}\right)^{n_s-1}
\left[y_1^{n_s}\vartheta T^2(ky_1){\cal P}_L(k\vartheta)+
\vartheta^{n_s}y_1T^2(k\vartheta){\cal P}_L(ky_1)\right]
\right.
\nonumber\\
&+&\left.
\frac{16}{50625}(1+z_{dec})^4\left(\frac{\Omega_{\gamma,0}}{\Omega_{m,0}}\right)^4y_1\vartheta{\cal P}_L(ky_1)
{\cal P}_L(k\vartheta)
\right].
\end{eqnarray}
where $\kappa_{\ell}\equiv1-\frac{1}{2}(\ell+\frac{1}{2})\left(\frac{k}{a_0H_0}\right)^{-1}$ and $\vartheta\equiv\sqrt{1-2yy_1+y_1^2}$.
In the numerical solutions in figure \ref{fig1} the adiabatic mode is determined by the best fit parameters of {\it WMAP 9} only
\cite{wmap9}: $A_s=\Delta^2_{\cal R}=2.41\times 10^{-9}$, $n_s=0.972$, $k_p=0.002$ Mpc$^{-1}$.
The optical depth at reionization $\tau_r=0.089$ with the corresponding redshift $z_r=10.6$.
The magnetic field energy density over photon energy density is given by 
$\frac{\rho_{B,0}}{\rho_{\gamma,0}}=9.54\times 10^{-8}\left(\frac{B}{\rm nG}\right)^2$.
The maximal undamped wave number of the magnetic field spectrum  is $k_m=286.91\left(\frac{B}{\rm nG}\right)^{-1}$ Mpc$^{-1}$.
The angular power spectrum of the secondary temperature anisotropies  induced by the vector mode at linear order
is given by
\begin{eqnarray}
C_{\ell}^{V,V}&=&\frac{\pi}{9}\frac{\ell+1}{(\ell+\frac{1}{2})^2}\left(\frac{\Omega_{\gamma,0}}{\Omega_{b,0}}\right)^2
\left(1-e^{-\tau_r}\right)^2(1+z_r)\int\frac{dk}{k}\left(\frac{k}{a_0H_0}\right)^4\kappa^2_{\ell}
e^{-32\left[\sqrt{1+z_r}\kappa_{\ell}-1\right]^2}
\nonumber\\
&\times&
\int_0^{\infty}dy_1\int_{-1}^1dy\frac{1+y^2}{y_1}
{\cal P}_{\langle\pi_B^{(\pm 1)*}\pi_B^{\pm 1}\rangle}(ky_1)
\left[\frac{4}{25}A_s\left(\frac{k}{k_p}\right)^{n_s-1}\vartheta^{n_s}T^2(k\vartheta)
\right.
\nonumber\\
&&\left.
+\frac{4}{225}(1+z_{dec})^2\left(\frac{\Omega_{\gamma,0}}{\Omega_{m,0}}\right)^2\vartheta
{\cal P}_L(k\vartheta)
\right].
\end{eqnarray}
The total secondary temperature anisotropies due to the  bulk motion of the scatteres is then given by 
\begin{eqnarray}
C_{\ell}^{V}=C_{\ell}^{V,S}+C_{\ell}^{V,V}
\end{eqnarray}
which is shown together with the indivual contributions sourced by the scalar and vector mode,
respectively, in figures \ref{fig1} and \ref{fig2}.

As can be appreciated from figure \ref{fig1} the contribution induced by the scalar mode is important on larger 
scales as compared to  that  which is sourced by the magnetic vector mode.
For a magnetic field of 3 nG a local maximum is observed at $\ell\sim 2\times 10^4$. This is shifted towards larger values of 
$\ell$ for smaller field strengths. The formation of a local maximum is due to the cut-off of the matter power spectrum of the magnetic mode
at the wave number corresponding to the magnetic Jeans scale.
Not taking into account this cut-off would actually lead to a monotonous  increase on these scales  in the angular power spectrum due to the magnetic field
contribution. The contribution due to the vector mode dominates on very small scales. For magnetic fields of strength 3 nG 
this happens for $\ell\sim 2\times 10^5$. For weaker magnetic fields the domain where the vector mode induced contribution is important is shifted to even larger values of $\ell$.

Current data from SPT and ACT do not constrain this contribution to  the secondary CMB anisotropies. However,
with future observations from {\it Atacama Large Millimeter/submillimeter Array}  (ALMA)\footnote{\tt https://almascience.nrao.edu/}
\cite{alma}, for example,  the interesting region between $10^4<\ell<10^6$ might be reached.
A similar strong peak on very small angular scales, $\ell>10^5$ is also predicted, e.g., 
by scattering of CMB photons within proto galactic clouds \cite{PJ} or
by the effects of massive black hole formation \cite{ABS}.
In comparison, secondary CMB anisotropies induced by the kinetic Sunyaev-Zeldovich effect in a patchy reionized universe 
leads typically to a plateau over a range $10^3<\ell<10^6$ at a lower amplitude \cite{vbs,ams}.
The contribution to the thermal Sunyaev-Zeldovich effect due to the presence of a primordial magnetic field 
has been studied in \cite{ts} \cite{sl2}.
It leads to a significant rise in the angular power spectrum of the temperature fluctuations on scales larger than 
those where the Ostriker-Vishniac effect is important.
In particular for magnetic fields which are not close to scale invariance observations for $\ell< 10^4$ constrain 
quite strongly the magnetic field parameters \cite{ts,sl2}.
Depending on details of the calculation of the matter perturbations limits are more \cite{ts} or less stringent \cite{sl2}.

\section{Conclusions}
\setcounter{equation}{0}

The secondary temperature anisotropies caused by the combination of inhomogeneous distribution and 
movement of electrons and generally ionized matter  along the line sight has been calculated in the 
presence of a stochastic magnetic field in the  post-decoupling, reionized universe. 
A simple model of reionziation has been used in which reionization is not assumed to be instanteneous but 
rather a  finite width of the re-scattering surface has been introduced by choosing the visibility function to be a gaussian.
As magnetic fields induce vector modes at linear order
there is an additional source term for the secondary anisotropies. Magnetic fields also have an important contribution
to the scalar mode on small scales. This can be seen in their effect on the total linear matter power spectrum
where there is a rise in power  on small scales due to the effect of the Lorentz term in the baryon velocity equation
\cite{sl,kmnbt}. 
Similarly, the Lorentz term is responsible for an increase on small scales of the angular power spectrum of the secondary CMB anisotropies 
calculated here. However, due to the cut-off at the magnetic Jeans scale for  matter density perturbations induced by the magnetic field
a local maximum results  at multipoles $\ell\stackrel{>}{\sim}10^4$. 
For even larger values of $\ell$ the contribution sourced by the 
magnetic vector mode becomes dominant.  For magnetic fields of strength 3 nG and negative spectral index 
this results in a rise in the angular power spectrum for $\ell>2\times 10^5$ corresponding to angular scales less than 3.2''.
These scales might be probed with ALMA in the future which might provide an interesting possibility to 
search for traces of large scale, cosmological magnetic fields.

\section{Acknowledgements}

I would like to thank an anonymous referee for very useful comments.
Furthermore I would like to thank the CP3 World Program for financial support and CP3-Origins at the University of Southern Denmark  for hospitality where part of this work was done.
Financial support by Spanish Science Ministry grants FIS2012-30926, FPA2009-10612 and CSD2007-00042 is gratefully acknowledged.


\bibliography{references}

\begin{thebibliography}{38}
\expandafter\ifx\csname natexlab\endcsname\relax\def\natexlab#1{#1}\fi
\expandafter\ifx\csname bibnamefont\endcsname\relax
  \def\bibnamefont#1{#1}\fi
\expandafter\ifx\csname bibfnamefont\endcsname\relax
  \def\bibfnamefont#1{#1}\fi
\expandafter\ifx\csname citenamefont\endcsname\relax
  \def\citenamefont#1{#1}\fi
\expandafter\ifx\csname url\endcsname\relax
  \def\url#1{\texttt{#1}}\fi
\expandafter\ifx\csname urlprefix\endcsname\relax\def\urlprefix{URL }\fi
\providecommand{\bibinfo}[2]{#2}
\providecommand{\eprint}[2][]{\url{#2}}

\bibitem[{\citenamefont{Reichardt et~al.}(2012)\citenamefont{Reichardt, Shaw,
  Zahn, Aird, Benson et~al.}}]{spt}
\bibinfo{author}{\bibfnamefont{C.}~\bibnamefont{Reichardt}},
  \bibinfo{author}{\bibfnamefont{L.}~\bibnamefont{Shaw}},
  \bibinfo{author}{\bibfnamefont{O.}~\bibnamefont{Zahn}},
  \bibinfo{author}{\bibfnamefont{K.}~\bibnamefont{Aird}},
  \bibinfo{author}{\bibfnamefont{B.}~\bibnamefont{Benson}},
  \bibnamefont{et~al.}, \bibinfo{journal}{Astrophys.J.}
  \textbf{\bibinfo{volume}{755}}, \bibinfo{pages}{70} (\bibinfo{year}{2012}),
  \eprint{1111.0932}.

\bibitem[{\citenamefont{Das et~al.}(2013)\citenamefont{Das, Louis, Nolta,
  Addison, Battistelli et~al.}}]{act}
\bibinfo{author}{\bibfnamefont{S.}~\bibnamefont{Das}},
  \bibinfo{author}{\bibfnamefont{T.}~\bibnamefont{Louis}},
  \bibinfo{author}{\bibfnamefont{M.~R.} \bibnamefont{Nolta}},
  \bibinfo{author}{\bibfnamefont{G.~E.} \bibnamefont{Addison}},
  \bibinfo{author}{\bibfnamefont{E.~S.} \bibnamefont{Battistelli}},
  \bibnamefont{et~al.} (\bibinfo{year}{2013}), \eprint{1301.1037}.

\bibitem[{\citenamefont{Ade et~al.}(2013)}]{planck}
\bibinfo{author}{\bibfnamefont{P.}~\bibnamefont{Ade}} \bibnamefont{et~al.}
  (\bibinfo{collaboration}{Planck Collaboration}) (\bibinfo{year}{2013}),
  \eprint{1303.5076}.

\bibitem[{\citenamefont{Hinshaw et~al.}(2013)}]{wmap9}
\bibinfo{author}{\bibfnamefont{G.}~\bibnamefont{Hinshaw}} \bibnamefont{et~al.}
  (\bibinfo{collaboration}{WMAP}), \bibinfo{journal}{Astrophys.J.Suppl.}
  \textbf{\bibinfo{volume}{208}}, \bibinfo{pages}{19} (\bibinfo{year}{2013}),
  \eprint{1212.5226}.

\bibitem[{\citenamefont{Reichardt et~al.}(2009)\citenamefont{Reichardt, Ade,
  Bock, Bond, Brevik et~al.}}]{acbar}
\bibinfo{author}{\bibfnamefont{C.}~\bibnamefont{Reichardt}},
  \bibinfo{author}{\bibfnamefont{P.}~\bibnamefont{Ade}},
  \bibinfo{author}{\bibfnamefont{J.}~\bibnamefont{Bock}},
  \bibinfo{author}{\bibfnamefont{J.~R.} \bibnamefont{Bond}},
  \bibinfo{author}{\bibfnamefont{J.}~\bibnamefont{Brevik}},
  \bibnamefont{et~al.}, \bibinfo{journal}{Astrophys.J.}
  \textbf{\bibinfo{volume}{694}}, \bibinfo{pages}{1200} (\bibinfo{year}{2009}),
  \eprint{0801.1491}.

\bibitem[{\citenamefont{Brown et~al.}(2009)}]{quad}
\bibinfo{author}{\bibfnamefont{M.}~\bibnamefont{Brown}} \bibnamefont{et~al.}
  (\bibinfo{collaboration}{QUaD collaboration}),
  \bibinfo{journal}{Astrophys.J.} \textbf{\bibinfo{volume}{705}},
  \bibinfo{pages}{978} (\bibinfo{year}{2009}), \eprint{0906.1003}.

\bibitem[{\citenamefont{Becker et~al.}(2001)}]{quasar}
\bibinfo{author}{\bibfnamefont{R.~H.} \bibnamefont{Becker}}
  \bibnamefont{et~al.} (\bibinfo{collaboration}{SDSS Collaboration}),
  \bibinfo{journal}{Astron.J.} \textbf{\bibinfo{volume}{122}},
  \bibinfo{pages}{2850} (\bibinfo{year}{2001}), \eprint{astro-ph/0108097}.

\bibitem[{\citenamefont{Vishniac}(1987)}]{vishniac}
\bibinfo{author}{\bibfnamefont{E.~T.} \bibnamefont{Vishniac}},
  \bibinfo{journal}{Astrophys.J.} \textbf{\bibinfo{volume}{322}},
  \bibinfo{pages}{597} (\bibinfo{year}{1987}).

\bibitem[{\citenamefont{Ostriker and Vishniac}(1986)}]{ov}
\bibinfo{author}{\bibfnamefont{J.}~\bibnamefont{Ostriker}} \bibnamefont{and}
  \bibinfo{author}{\bibfnamefont{E.}~\bibnamefont{Vishniac}},
  \bibinfo{journal}{Astrophys.J.} \textbf{\bibinfo{volume}{306}},
  \bibinfo{pages}{L51} (\bibinfo{year}{1986}).

\bibitem[{\citenamefont{Jaffe and Kamionkowski}(1998)}]{jk}
\bibinfo{author}{\bibfnamefont{A.~H.} \bibnamefont{Jaffe}} \bibnamefont{and}
  \bibinfo{author}{\bibfnamefont{M.}~\bibnamefont{Kamionkowski}},
  \bibinfo{journal}{Phys.Rev.} \textbf{\bibinfo{volume}{D58}},
  \bibinfo{pages}{043001} (\bibinfo{year}{1998}), \eprint{astro-ph/9801022}.

\bibitem[{\citenamefont{Hu}(2000)}]{hu}
\bibinfo{author}{\bibfnamefont{W.}~\bibnamefont{Hu}},
  \bibinfo{journal}{Astrophys.J.} \textbf{\bibinfo{volume}{529}},
  \bibinfo{pages}{12} (\bibinfo{year}{2000}), \eprint{astro-ph/9907103}.

\bibitem[{\citenamefont{Hu and White}(1996)}]{hw1}
\bibinfo{author}{\bibfnamefont{W.}~\bibnamefont{Hu}} \bibnamefont{and}
  \bibinfo{author}{\bibfnamefont{M.~J.} \bibnamefont{White}},
  \bibinfo{journal}{Astron.Astrophys.} \textbf{\bibinfo{volume}{315}},
  \bibinfo{pages}{33} (\bibinfo{year}{1996}), \eprint{astro-ph/9507060}.

\bibitem[{\citenamefont{Hu et~al.}(1994)\citenamefont{Hu, Scott, and
  Silk}}]{HuSS}
\bibinfo{author}{\bibfnamefont{W.}~\bibnamefont{Hu}},
  \bibinfo{author}{\bibfnamefont{D.}~\bibnamefont{Scott}}, \bibnamefont{and}
  \bibinfo{author}{\bibfnamefont{J.}~\bibnamefont{Silk}},
  \bibinfo{journal}{Phys.Rev.} \textbf{\bibinfo{volume}{D49}},
  \bibinfo{pages}{648} (\bibinfo{year}{1994}), \eprint{astro-ph/9305038}.

\bibitem[{\citenamefont{Dodelson and Jubas}(1995)}]{dj}
\bibinfo{author}{\bibfnamefont{S.}~\bibnamefont{Dodelson}} \bibnamefont{and}
  \bibinfo{author}{\bibfnamefont{J.~M.} \bibnamefont{Jubas}},
  \bibinfo{journal}{Astrophys.J.} \textbf{\bibinfo{volume}{439}},
  \bibinfo{pages}{503} (\bibinfo{year}{1995}), \eprint{astro-ph/9308019}.

\bibitem[{\citenamefont{Mack et~al.}(2002)\citenamefont{Mack, Kahniashvili, and
  Kosowsky}}]{mkk}
\bibinfo{author}{\bibfnamefont{A.}~\bibnamefont{Mack}},
  \bibinfo{author}{\bibfnamefont{T.}~\bibnamefont{Kahniashvili}},
  \bibnamefont{and} \bibinfo{author}{\bibfnamefont{A.}~\bibnamefont{Kosowsky}},
  \bibinfo{journal}{Phys.Rev.} \textbf{\bibinfo{volume}{D65}},
  \bibinfo{pages}{123004} (\bibinfo{year}{2002}), \eprint{astro-ph/0105504}.

\bibitem[{\citenamefont{Yamazaki et~al.}(2008)\citenamefont{Yamazaki, Ichiki,
  Kajino, and Mathews}}]{yikm}
\bibinfo{author}{\bibfnamefont{D.~G.} \bibnamefont{Yamazaki}},
  \bibinfo{author}{\bibfnamefont{K.}~\bibnamefont{Ichiki}},
  \bibinfo{author}{\bibfnamefont{T.}~\bibnamefont{Kajino}}, \bibnamefont{and}
  \bibinfo{author}{\bibfnamefont{G.~J.} \bibnamefont{Mathews}},
  \bibinfo{journal}{Phys.Rev.} \textbf{\bibinfo{volume}{D77}},
  \bibinfo{pages}{043005} (\bibinfo{year}{2008}), \eprint{0801.2572}.

\bibitem[{\citenamefont{Paoletti et~al.}(2009)\citenamefont{Paoletti, Finelli,
  and Paci}}]{pfp}
\bibinfo{author}{\bibfnamefont{D.}~\bibnamefont{Paoletti}},
  \bibinfo{author}{\bibfnamefont{F.}~\bibnamefont{Finelli}}, \bibnamefont{and}
  \bibinfo{author}{\bibfnamefont{F.}~\bibnamefont{Paci}},
  \bibinfo{journal}{Mon.Not.Roy.Astron.Soc.} \textbf{\bibinfo{volume}{396}},
  \bibinfo{pages}{523} (\bibinfo{year}{2009}), \eprint{0811.0230}.

\bibitem[{\citenamefont{Shaw and Lewis}(2010)}]{sl}
\bibinfo{author}{\bibfnamefont{J.~R.} \bibnamefont{Shaw}} \bibnamefont{and}
  \bibinfo{author}{\bibfnamefont{A.}~\bibnamefont{Lewis}},
  \bibinfo{journal}{Phys.Rev.} \textbf{\bibinfo{volume}{D81}},
  \bibinfo{pages}{043517} (\bibinfo{year}{2010}), \eprint{0911.2714}.

\bibitem[{\citenamefont{Kunze}(2012{\natexlab{a}})}]{kk1}
\bibinfo{author}{\bibfnamefont{K.~E.} \bibnamefont{Kunze}},
  \bibinfo{journal}{Phys.Rev.} \textbf{\bibinfo{volume}{D85}},
  \bibinfo{pages}{083004} (\bibinfo{year}{2012}{\natexlab{a}}),
  \eprint{1112.4797}.

\bibitem[{\citenamefont{Kodama and Sasaki}(1984)}]{ks}
\bibinfo{author}{\bibfnamefont{H.}~\bibnamefont{Kodama}} \bibnamefont{and}
  \bibinfo{author}{\bibfnamefont{M.}~\bibnamefont{Sasaki}},
  \bibinfo{journal}{Prog.Theor.Phys.Suppl.} \textbf{\bibinfo{volume}{78}},
  \bibinfo{pages}{1} (\bibinfo{year}{1984}).

\bibitem[{\citenamefont{Hu and White}(1997)}]{hw}
\bibinfo{author}{\bibfnamefont{W.}~\bibnamefont{Hu}} \bibnamefont{and}
  \bibinfo{author}{\bibfnamefont{M.~J.} \bibnamefont{White}},
  \bibinfo{journal}{Phys.Rev.} \textbf{\bibinfo{volume}{D56}},
  \bibinfo{pages}{596} (\bibinfo{year}{1997}), \eprint{astro-ph/9702170}.

\bibitem[{\citenamefont{Subramanian and Barrow}(1998)}]{sb}
\bibinfo{author}{\bibfnamefont{K.}~\bibnamefont{Subramanian}} \bibnamefont{and}
  \bibinfo{author}{\bibfnamefont{J.~D.} \bibnamefont{Barrow}},
  \bibinfo{journal}{Phys.Rev.} \textbf{\bibinfo{volume}{D58}},
  \bibinfo{pages}{083502} (\bibinfo{year}{1998}), \eprint{astro-ph/9712083}.

\bibitem[{\citenamefont{Jedamzik et~al.}(1998)\citenamefont{Jedamzik,
  Katalinic, and Olinto}}]{jko}
\bibinfo{author}{\bibfnamefont{K.}~\bibnamefont{Jedamzik}},
  \bibinfo{author}{\bibfnamefont{V.}~\bibnamefont{Katalinic}},
  \bibnamefont{and} \bibinfo{author}{\bibfnamefont{A.~V.}
  \bibnamefont{Olinto}}, \bibinfo{journal}{Phys.Rev.}
  \textbf{\bibinfo{volume}{D57}}, \bibinfo{pages}{3264} (\bibinfo{year}{1998}),
  \eprint{astro-ph/9606080}.

\bibitem[{\citenamefont{Kunze and Komatsu}(2014)}]{kuko}
\bibinfo{author}{\bibfnamefont{K.~E.} \bibnamefont{Kunze}} \bibnamefont{and}
  \bibinfo{author}{\bibfnamefont{E.}~\bibnamefont{Komatsu}},
  \bibinfo{journal}{JCAP} \textbf{\bibinfo{volume}{01}}, \bibinfo{pages}{009}
  (\bibinfo{year}{2014}), \eprint{1309.7994}.

\bibitem[{\citenamefont{Sethi and Subramanian}(2005)}]{sesu}
\bibinfo{author}{\bibfnamefont{S.~K.} \bibnamefont{Sethi}} \bibnamefont{and}
  \bibinfo{author}{\bibfnamefont{K.}~\bibnamefont{Subramanian}},
  \bibinfo{journal}{Mon.Not.Roy.Astron.Soc.} \textbf{\bibinfo{volume}{356}},
  \bibinfo{pages}{778} (\bibinfo{year}{2005}), \eprint{astro-ph/0405413}.

\bibitem[{\citenamefont{Kim et~al.}(1996)\citenamefont{Kim, Olinto, and
  Rosner}}]{kor}
\bibinfo{author}{\bibfnamefont{E.-j.} \bibnamefont{Kim}},
  \bibinfo{author}{\bibfnamefont{A.}~\bibnamefont{Olinto}}, \bibnamefont{and}
  \bibinfo{author}{\bibfnamefont{R.}~\bibnamefont{Rosner}},
  \bibinfo{journal}{Astrophys.J.} \textbf{\bibinfo{volume}{468}},
  \bibinfo{pages}{28} (\bibinfo{year}{1996}), \eprint{astro-ph/9412070}.

\bibitem[{\citenamefont{Peter and Uzan}(2009)}]{pu}
\bibinfo{author}{\bibfnamefont{P.}~\bibnamefont{Peter}} \bibnamefont{and}
  \bibinfo{author}{\bibfnamefont{J.-P.} \bibnamefont{Uzan}},
  \emph{\bibinfo{title}{Primordial Cosmology}} (\bibinfo{publisher}{Oxford
  University Press}, \bibinfo{year}{2009}).

\bibitem[{\citenamefont{Bardeen et~al.}(1986)\citenamefont{Bardeen, Bond,
  Kaiser, and Szalay}}]{bbks}
\bibinfo{author}{\bibfnamefont{J.~M.} \bibnamefont{Bardeen}},
  \bibinfo{author}{\bibfnamefont{J.}~\bibnamefont{Bond}},
  \bibinfo{author}{\bibfnamefont{N.}~\bibnamefont{Kaiser}}, \bibnamefont{and}
  \bibinfo{author}{\bibfnamefont{A.}~\bibnamefont{Szalay}},
  \bibinfo{journal}{Astrophys.J.} \textbf{\bibinfo{volume}{304}},
  \bibinfo{pages}{15} (\bibinfo{year}{1986}).

\bibitem[{\citenamefont{Kunze}(2012{\natexlab{b}})}]{kk2}
\bibinfo{author}{\bibfnamefont{K.~E.} \bibnamefont{Kunze}},
  \bibinfo{journal}{Phys.Rev.} \textbf{\bibinfo{volume}{D85}},
  \bibinfo{pages}{083004} (\bibinfo{year}{2012}{\natexlab{b}}),
  \eprint{1112.4797}.

\bibitem[{\citenamefont{Gopal and Sethi}(2005)}]{gs}
\bibinfo{author}{\bibfnamefont{R.}~\bibnamefont{Gopal}} \bibnamefont{and}
  \bibinfo{author}{\bibfnamefont{S.~K.} \bibnamefont{Sethi}},
  \bibinfo{journal}{Phys.Rev.} \textbf{\bibinfo{volume}{D72}},
  \bibinfo{pages}{103003} (\bibinfo{year}{2005}), \eprint{astro-ph/0506642}.

\bibitem[{\citenamefont{Di~Francesco et~al.}(2013)\citenamefont{Di~Francesco,
  Johnstone, Matthews, Bartel, Bronfman et~al.}}]{alma}
\bibinfo{author}{\bibfnamefont{J.}~\bibnamefont{Di~Francesco}},
  \bibinfo{author}{\bibfnamefont{D.}~\bibnamefont{Johnstone}},
  \bibinfo{author}{\bibfnamefont{B.}~\bibnamefont{Matthews}},
  \bibinfo{author}{\bibfnamefont{N.}~\bibnamefont{Bartel}},
  \bibinfo{author}{\bibfnamefont{L.}~\bibnamefont{Bronfman}},
  \bibnamefont{et~al.} (\bibinfo{year}{2013}), \eprint{1310.1604}.

\bibitem[{\citenamefont{Peebles and Juszkiewicz}(1998)}]{PJ}
\bibinfo{author}{\bibfnamefont{P.}~\bibnamefont{Peebles}} \bibnamefont{and}
  \bibinfo{author}{\bibfnamefont{R.}~\bibnamefont{Juszkiewicz}},
  \bibinfo{journal}{Astrophys.J.} \textbf{\bibinfo{volume}{509}},
  \bibinfo{pages}{483} (\bibinfo{year}{1998}), \eprint{astro-ph/9804260}.

\bibitem[{\citenamefont{Aghanim et~al.}(2000)\citenamefont{Aghanim, Balland,
  and Silk}}]{ABS}
\bibinfo{author}{\bibfnamefont{N.}~\bibnamefont{Aghanim}},
  \bibinfo{author}{\bibfnamefont{C.}~\bibnamefont{Balland}}, \bibnamefont{and}
  \bibinfo{author}{\bibfnamefont{J.}~\bibnamefont{Silk}},
  \bibinfo{journal}{Astron.Astrophys.} \textbf{\bibinfo{volume}{357}},
  \bibinfo{pages}{1} (\bibinfo{year}{2000}), \eprint{astro-ph/0003254}.

\bibitem[{\citenamefont{Valageas et~al.}(2001)\citenamefont{Valageas, Balbi,
  and Silk}}]{vbs}
\bibinfo{author}{\bibfnamefont{P.}~\bibnamefont{Valageas}},
  \bibinfo{author}{\bibfnamefont{A.}~\bibnamefont{Balbi}}, \bibnamefont{and}
  \bibinfo{author}{\bibfnamefont{J.}~\bibnamefont{Silk}},
  \bibinfo{journal}{Astron.Astrophys.} \textbf{\bibinfo{volume}{367}},
  \bibinfo{pages}{1} (\bibinfo{year}{2001}), \eprint{astro-ph/0009040}.

\bibitem[{\citenamefont{Aghanim et~al.}(2008)\citenamefont{Aghanim, Majumdar,
  and Silk}}]{ams}
\bibinfo{author}{\bibfnamefont{N.}~\bibnamefont{Aghanim}},
  \bibinfo{author}{\bibfnamefont{S.}~\bibnamefont{Majumdar}}, \bibnamefont{and}
  \bibinfo{author}{\bibfnamefont{J.}~\bibnamefont{Silk}},
  \bibinfo{journal}{Rept.Prog.Phys.} \textbf{\bibinfo{volume}{71}},
  \bibinfo{pages}{066902} (\bibinfo{year}{2008}), \eprint{0711.0518}.

\bibitem[{\citenamefont{Tashiro and Sugiyama}(2009)}]{ts}
\bibinfo{author}{\bibfnamefont{H.}~\bibnamefont{Tashiro}} \bibnamefont{and}
  \bibinfo{author}{\bibfnamefont{N.}~\bibnamefont{Sugiyama}},
  \bibinfo{journal}{Mon.Not.Roy.Astron.Soc.} \textbf{\bibinfo{volume}{411}},
  \bibinfo{pages}{1284} (\bibinfo{year}{2009}), \eprint{0908.0113}.

\bibitem[{\citenamefont{Shaw and Lewis}(2012)}]{sl2}
\bibinfo{author}{\bibfnamefont{J.~R.} \bibnamefont{Shaw}} \bibnamefont{and}
  \bibinfo{author}{\bibfnamefont{A.}~\bibnamefont{Lewis}},
  \bibinfo{journal}{Phys.Rev.} \textbf{\bibinfo{volume}{D86}},
  \bibinfo{pages}{043510} (\bibinfo{year}{2012}), \eprint{1006.4242}.

\bibitem[{\citenamefont{Kahniashvili et~al.}(2013)\citenamefont{Kahniashvili,
  Maravin, Natarajan, Battaglia, and Tevzadze}}]{kmnbt}
\bibinfo{author}{\bibfnamefont{T.}~\bibnamefont{Kahniashvili}},
  \bibinfo{author}{\bibfnamefont{Y.}~\bibnamefont{Maravin}},
  \bibinfo{author}{\bibfnamefont{A.}~\bibnamefont{Natarajan}},
  \bibinfo{author}{\bibfnamefont{N.}~\bibnamefont{Battaglia}},
  \bibnamefont{and} \bibinfo{author}{\bibfnamefont{A.~G.}
  \bibnamefont{Tevzadze}}, \bibinfo{journal}{Astrophys.J.}
  \textbf{\bibinfo{volume}{770}}, \bibinfo{pages}{47} (\bibinfo{year}{2013}),
  \eprint{1211.2769}.

\end{thebibliography}

\bibliographystyle{apsrev}

\end{document}